\documentclass[apj]{emulateapj}
\usepackage{apjfonts}
\usepackage{graphicx}
\usepackage{amsmath}

\slugcomment{ApJ, in press}

\begin{document}

\vspace{1mm}

\shortauthors{Bartoli et al.} 

\shorttitle{ARGO-YBJ observation of MGRO J1908+06} 


\title {Observation of the TeV gamma-ray source MGRO J1908+06 with ARGO-YBJ}



\author{B.~Bartoli\altaffilmark{1,2},
 P.~Bernardini\altaffilmark{3,4},
 X.J.~Bi\altaffilmark{5},
 C.~Bleve\altaffilmark{3,4},
 I.~Bolognino\altaffilmark{6,7},
 P.~Branchini\altaffilmark{8},
 A.~Budano\altaffilmark{8},
 A.K.~Calabrese Melcarne\altaffilmark{9},
 P.~Camarri\altaffilmark{10,11},
 Z.~Cao\altaffilmark{5}
% ...
\\ (The Argo-YBJ Collaboration)}


  \affil{\altaffilmark{1} Dipartimento di Fisica dell'Universit\`a di Napoli
                  ``Federico II'', Complesso Universitario di Monte 
                  Sant'Angelo, via Cinthia, 80126 Napoli, Italy.}
  \affil{\altaffilmark{2} Istituto Nazionale di Fisica Nucleare, Sezione di
                  Napoli, Complesso Universitario di Monte
                  Sant'Angelo, via Cinthia, 80126 Napoli, Italy.}
  \affil{\altaffilmark{3} Dipartimento di Fisica dell'Universit\`a del Salento,
                  via per Arnesano, 73100 Lecce, Italy.}
  \affil{\altaffilmark{4} Istituto Nazionale di Fisica Nucleare, Sezione di
                  Lecce, via per Arnesano, 73100 Lecce, Italy.}
  \affil{\altaffilmark{5} Key Laboratory of Particle Astrophysics, Institute 
                  of High Energy Physics, Chinese Academy of Sciences,
                  P.O. Box 918, 100049 Beijing, P.R. China.}
  \affil{\altaffilmark{6} Dipartimento di Fisica Nucleare e Teorica 
                  dell'Universit\`a di Pavia, via Bassi 6,
                  27100 Pavia, Italy.}
  \affil{\altaffilmark{7} Istituto Nazionale di Fisica Nucleare, Sezione di Pavia, 
                  via Bassi 6, 27100 Pavia, Italy.}
  \affil{\altaffilmark{8} Istituto Nazionale di Fisica Nucleare, Sezione di
                  Roma Tre, via della Vasca Navale 84, 00146 Roma, Italy.}
  \affil{\altaffilmark{9} Istituto Nazionale di Fisica Nucleare - CNAF, Viale 
                  Berti-Pichat 6/2, 40127 Bologna, Italy.}
  \affil{\altaffilmark{10} Dipartimento di Fisica dell'Universit\`a di Roma ``Tor 									 Vergata'', 
                   via della Ricerca Scientifica 1, 00133 Roma, Italy.}
  \affil{\altaffilmark{11} Istituto Nazionale di Fisica Nucleare, Sezione di
                   Roma Tor Vergata, via della Ricerca Scientifica 1, 
                   00133 Roma, Italy.}
  \affil{\altaffilmark{12} Osservatorio Astrofisico di Torino 
                   dell'Istituto Nazionale di Astrofisica, corso Fiume 4, 
                   10133 Torino, Italy.}
  \affil{\altaffilmark{13} Istituto Nazionale di Fisica Nucleare,
                   Sezione di Torino, via P. Giuria 1, 10125 Torino, Italy.}
  \affil{\altaffilmark{14} Dipartimento di Fisica dell'Universit\`a ``Roma Tre'', 
                   via della Vasca Navale 84, 00146 Roma, Italy.}
  \affil{\altaffilmark{15} Tibet University, 850000 Lhasa, Xizang, P.R. China.}
  \affil{\altaffilmark{16} Hebei Normal University, Shijiazhuang 050016, 
                   Hebei, P.R. China.}
  \affil{\altaffilmark{17} Yunnan University, 2 North Cuihu Rd., 650091 Kunming, 
                   Yunnan, P.R. China.}
  \affil{\altaffilmark{18} Universit\`a degli Studi di Palermo, Dipartimento di Fisica 
                   e Tecnologie Relative, Viale delle Scienze, Edificio 18, 
                   90128 Palermo, Italy.}
  \affil{\altaffilmark{19} Istituto Nazionale di Fisica Nucleare, Sezione di Catania, 
                   Viale A. Doria 6, 95125 Catania, Italy.}
  \affil{\altaffilmark{20} Dipartimento di Fisica dell'Universit\`a di Torino,
                   via P. Giuria 1, 10125 Torino, Italy.}
  \affil{\altaffilmark{21} Shandong University, 250100 Jinan, Shandong, P.R. China.}
  \affil{\altaffilmark{22} Southwest Jiaotong University, 610031 Chengdu, 
                   Sichuan, P.R. China.}
  \affil{\altaffilmark{23} Dipartimento di Ingegneria dell'Innovazione,  
                   Universit\`a del Salento, 73100 Lecce, Italy.}
  \affil{\altaffilmark{24} Istituto di Astrofisica Spaziale e Fisica Cosmica 
                   dell'Istituto Nazionale di Astrofisica, 
                   via La Malfa 153, 90146 Palermo, Italy.}

\email{vernetto@to.infn.it}

\begin{abstract}

The extended gamma ray source MGRO J1908+06, discovered by the Milagro
air shower detector in 2007, has been observed for $\sim$4 years by
the ARGO-YBJ experiment at TeV energies, with a statistical significance
of 6.2 standard deviations.
The peak of the signal is found at a position consistent with the pulsar
PSR J1907+0602.
Parametrizing the source
shape with a two-dimensional Gauss function we estimate an 
extension $\sigma_{ext}$ = 0.49$^{\circ}$$\pm$0.22$^{\circ}$, consistent with
a previous measurement by the Cherenkov Array H.E.S.S..  
The observed energy spectrum is 
dN/dE = 6.1$\pm$1.4 $\times$ 10$^{-13}$ (E/4 TeV)$^{-2.54\pm0.36}$ 
photons cm$^{-2}$ s$^{-1}$ TeV$^{-1}$,
in the energy range $\sim$1-20 TeV.
The measured gamma ray flux is consistent with the
results of the Milagro detector, but is $\sim$2-3 times larger than 
the flux previously derived by H.E.S.S. at energies of a few TeV. 
The continuity of the Milagro and ARGO-YBJ observations and the stable excess
rate observed by ARGO-YBJ along 4 years of data taking support the 
identification of MGRO J1908+06 as the steady powerful TeV 
pulsar wind nebula of PSR J1907+0602, with an integrated luminosity above
1 TeV $\sim$1.8 times the Crab Nebula luminosity.

\end{abstract}



\section{Introduction}

The Galactic gamma ray source MGRO J1908+06 was discovered by the Milagro 
air shower detector in a survey of the Galactic plane
at a median energy of $\sim 20$ TeV \citep{mila07}.
The data were consistent both with a point source and with an extended 
source of diameter $<$2.6$^{\circ}$.
Assuming a spectrum $\propto$E$^{-2.3}$,
the measured flux at the median energy of 20 TeV is
8.8$\pm$2.4 $\times$ 10$^{-15}$ photons cm$^{-2}$ s$^{-1}$ TeV$^{-1}$.

A marginal detection of a source consistent with the position of MGRO J1908+06
was already reported by the Tibet AS-$\gamma$ array \citep{Zha03},
but not confirmed in a more recent paper \citep{Ane10}.

The source was later observed 
by the H.E.S.S. \citep{hess09} and VERITAS 
\citep{ward08} Cherenkov telescopes.
In particular, H.E.S.S  detected an extended source (HESS J1908+063)
at energies above 300 GeV 
($\sim$11 standard deviations of statistical significance)
positionally consistent with MGRO J1908+06.
The measured source extension, evaluated assuming a symmetrical 
two-dimensional Gaussian shape, was 
$\sigma_{ext}$ = 0.34$^{\circ}$$_{-0.03}^{+0.04}$. 

H.E.S.S. reported a power law differential energy spectrum with a
photon index of 
2.10 $\pm$ 0.07$_{stat}$ $\pm$ 0.2$_{sys}$ in the energy range 0.3-20 TeV, 
and a flux at 1 TeV of (4.14 $\pm$ 0.32 $_{stat}$ $\pm$ 0.83$_{sys}$)
$\times$ 10$^{-12}$ photons cm$^{-2}$ s$^{-1}$ TeV$^{-1}$.
The integrated flux above 1 TeV is 17$\%$ that of the Crab Nebula.

After the release of the Bright Source List by the Fermi collaboration
\citep{ferm09}, Milagro reported the association of
MGRO J1908+06 to the LAT pulsar 0FGL J1907.5+0602 (later renamed 
PSR J1907+0602), pulsating with a period of 106.6 ms \citep{mila09}.
The peak of the Milagro emission was 0.3$^{\circ}$ off the pulsar, but
consistent with the pulsar location within the 
measurement error (0.27$^{\circ}$). 
Assuming a spectrum $\propto$E$^{-2.6}$, Milagro reported a flux of
116.7$\pm$15.8 $\times$ 10$^{-17}$ photons cm$^{-2}$ s$^{-1}$ TeV$^{-1}$, 
at the median energy of 35 TeV.

The association of MGRO J1908+06 with PSR J1907+0602 
was also supported in \citep{fermi10}, where a 
multiwavelength study of the pulsar and the surrounding region 
has been performed with radio, X-ray and Fermi gamma ray data. 
Because of the small angular distance between the pulsar 
and the centroid of the H.E.S.S. 
extended source, the authors argue that the latter
is plausibly the Wind Nebula of the pulsar.

Performing an off-pulse measurement, Fermi set an upper limit to the 
HESS J1908+063 flux in the energy region 0.1-25 GeV, 
suggesting that the spectrum
has a low-energy turnover between 20 GeV and 300 GeV.
With radio and X-ray data, a lower limit to the pulsar distance was set 
to $\sim$3.2 kpc, deriving for the nebula a physical size
$\ge$40 pc.

Later, Milagro evaluated
the energy spectrum of the source in the 2-100 TeV 
region, reporting a hard power law spectrum with an exponential
cutoff \citep{mila09b}.
The best fit obtained is  
dN/dE = 0.62 $\times$ 10$^{-11}$ E$^{-1.50}$ exp(-E/14.1) photons
cm$^{-2}$s$^{-1}$ TeV$^{-1}$, where E is the energy in TeV.
This flux is in disagreement with that given by H.E.S.S. at a level
of 2-3 standard deviations, being about a factor 3 higher at 10 TeV.
The authors suggest that the discrepancy can be simply due  
to a statistical fluctuation, or to the fact that
Milagro, given its relatively poor angular resolution,
integrates the signal over a larger solid angle compared with H.E.S.S., 
and likely detects more of the diffuse lateral tails of the 
extended source.

In this work we report on the 
observation of MGRO J1908+06 with the ARGO-YBJ detector
performed during the years 2007-2011.
After a brief description of the detector and a detailed presentation of
the data analysis technique, we 
report our results concerning the extension and the energy 
spectrum of the source.

\section {The ARGO-YBJ experiment}

The ARGO-YBJ detector is located at the Yangbajing Cosmic Ray Laboratory 
(Tibet, China) at an altitude of 4300 m above sea level.
It consists of a 
$\sim$74$\times$ 78 m$^2$ carpet made of a single layer of Resistive
Plate Chambers (RPCs) with $\sim$92$\%$ of active area, sorrounded
by a partially instrumented ($\sim$20$\%$) area up to
$\sim$100$\times$110 m$^2$. 
The apparatus has a modular structure,
the basic data acquisition element being a cluster (5.7$\times$7.6
m$^2$), made of 12 RPCs (2.8$\times$1.25 m$^2$). 
The RPCs are operated in streamer mode by using a gas mixture (Ar 15$\%$,
Isobutane 10$\%$, TetraFluoroEthane 75$\%$) suitable 
for high altitude operation.

Each RPC is read by 80 strips of 6.75$\times$61.8 cm$^2$ (the
spatial pixels), logically organized in 10 independent pads of
55.6$\times$61.8 cm$^2$ which are individually acquired and
represent the time pixels of the detector \citep{Aie06}.  
In addition, in order to extend the dynamical range 
up to PeV energies, each RPC is equipped with two large size pads (139 $\times$ 123 cm$^2$) to collect the total charge developed by the particle hitting
the detector \citep{iaco09}.
The full experiment is made of 153 clusters for a total active surface 
of $\sim$6600m$^2$. 
 
ARGO-YBJ operates in two independent
acquisition modes: the {\em shower mode} and the {\em scaler mode}
\citep{Aie08}. In this analysis we refer to the data recorded
from the digital read-out in shower mode.
In this mode, an electronic logic has been implemented
to build an inclusive trigger, based on a time correlation 
between the pad signals, depending on their relative distances. 
In this way, all the shower events giving
a number of fired pads 
N$_{pad}\ge$ N$_{trig}$ in the central carpet in a time window 
of 420 ns generate the trigger.
This trigger can work with high efficiency down to N$_{trig}$=20,
keeping negligible the rate of random coincidences \citep{Alo04}.

The time of each
fired pad in a window of 2 $\mu$sec around the trigger time and its location are recorded
and used to reconstruct the position of the shower
core and the arrival direction of the primary particle.

In order to perform the time calibration of the 18360 pads, a
software procedure has been developed,
based on the Characteristic Plane method  \citep{He07}
which using the secondary particles of large vertical 
showers as calibration beams,
iteratively reduces the differences between the 
measured times and the temporal fit of the shower front  \citep{Aie09}.

The full detector is in stable data taking since 2007 November
with the trigger condition N$_{trig}$=20 and a duty cycle $\sim
86\%$. The trigger rate is $\sim$3.5 kHz with a dead time of 4$\%$.

\section{Detector performance}

The angular resolution and the pointing accuracy of the detector
have been evaluated by using the Moon shadow, i.e. the deficit of
cosmic rays in the Moon direction, observed
by ARGO-YBJ with a statistical significance 
of $\sim$9 standard deviations per month.
The shape of the shadow
provides a measurement of the detector Point Spread Function
(PSF), and its position allows the individuation of possible pointing biases.

The data have been compared with the results of a Monte Carlo simulation 
which describes
the propagation of cosmic rays in the Earth magnetic fields, 
the shower development in the atmosphere by using the 
CORSIKA code \citep{Hec98}, and the detector response with a code based
on the GEANT package \citep{Gea93}.
The PSF measured with the cosmic rays has been found 
in excellent agreement with the Monte Carlo
evaluation, confirming the reliability of the simulation procedure \citep{DiS11}. 

The angular resolution for gamma rays is
evaluated by simulating the events
from a gamma ray source with a given spectrum and
daily path in the sky.
It results smaller by $\sim$30-40$\%$ compared with the
angular resolution for cosmic rays, due
to the better defined time profile of the showers. In general, the PSF
for gamma rays can be described by the sum of
two Gaussian distributions. 
For a Crab-like source, the radius of the opening angle 
which optimizes the signal-to-background ratio 
for events with N$_{pad}\geq$60 (300)
is 0.86$^{\circ}$ (0.44$^{\circ}$)
and contains $\sim$50$\%$ of the signal.

The Moon Shadow has also been used to check the absolute energy calibration
of the detector, by studying the westward shift of the shadow
due to the geomagnetic field.
The observed displacement as a function of the event multiplicity N$_{pad}$ 
is in excellent agreement with the results of the Monte Carlo simulation.
From this analysis the total absolute energy scale error,
including systematic effects, is estimated to be less than 13$\%$ 
\citep{DiS11}.

\begin{figure}
    \includegraphics[height=0.33\textheight]{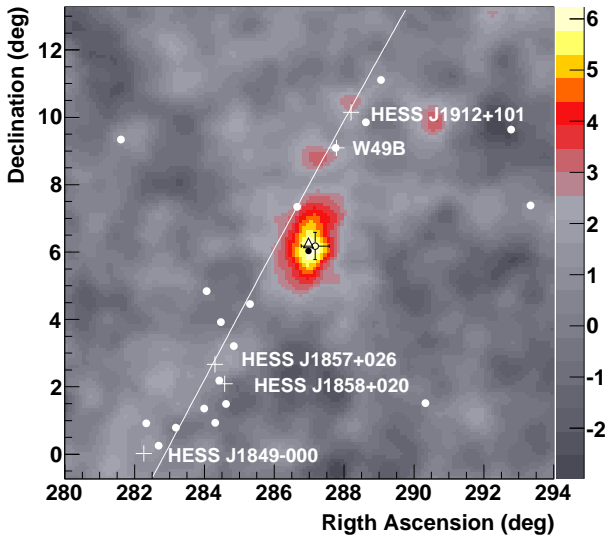}
    \caption{PSF-smoothed significance map of the  MGRO J1908+06 
region obtained by ARGO-YBJ, 
for events with  N$_{pad} \ge$60.
Open circle: position of the center of MGRO J1908+06, as measured by Milagro.
The error bars give the linear sum of the statistical and systematic errors.
Open triangle: centroid of HESS J1908+063.
Black filled circle: Fermi pulsar PSR J1907+0602.
White filled circles: Fermi gamma ray sources, according to the 
2nd Fermi Catalogue \citep{fermicat}. White crosses:
TeV sources detected by H.E.S.S. in the same region. The white line represents 
the Galactic plane.
}
\vspace{15pt}
\end{figure}


\section{Data analysis and results}

At the ARGO-YBJ site MGRO J1908+06 culminates at the 
zenith angle of 24$^{\circ}$ and is visible for 5.38 hours 
per day with a zenith angle less than 45$^{\circ}$.
The dataset used in this analysis
refers to the period from November 2007 to December 2011 and 
contains all the showers 
with zenith angle less than 45$^{\circ}$ and N$_{pad} \geq$20.
The total on-source time is 6867 hours. 

To study the gamma ray emission from a source, 
a 16$^{\circ}\times$16$^{\circ}$ 
sky map in celestial coordinates (right ascension and declination) with
0.1$^{\circ}\times$0.1$^{\circ}$ 
bin size, centered on the source
position, is filled with the detected events. 

In order to extract the excess of gamma rays, the cosmic ray
background has to be estimated and subtracted.

The {\em time swapping} method \citep{Ale92} is used to evaluate the
background: for each detected
event, $n$ "fake" events (with $n$ = 10) are generated by replacing the original
arrival time with new ones, randomly selected from an event buffer which
spans a time T of data taking. 
Changing the time, the fake events maintain the same declination of 
the original event,
but have a different right ascension. With these events a new sky
map (background map) is built, with a statistics $n$ times larger
than the ``true'' event map in order to reduce the fluctuations. 
To avoid the inclusion of the source events in the background evaluation,
the showers inside a circular region around the source (with a radius 
related to the PSF and depending on N$_{pad}$) are excluded 
from the time swapping procedure. A correction of the number of swaps 
is made to take into account 
the rejected events in the source region \citep{Fle04}.
The value of the swapping time T is $\sim$ 3 hours, in order to minimize the
systematic effects due to the environmental parameter variations.

In order to extract the source signal the maps are smoothed according
to the detector PSF, determined by Monte Carlo simulations for different
N$_{pad}$ intervals.
Finally, the smoothed background map is subtracted to the
smoothed event map, obtaining the "excess map",
where for every bin the statistical significance S of the excess is
given by:

\begin{displaymath}
S = (N_{on}-N_{off})/\sqrt{\delta N_{on}^2+ \delta N_{off}^2} 
\end{displaymath}

\noindent with $N_{on}$ = $\Sigma_i$ $N_i$ $w_i$  and
$N_{off}$ = $\Sigma_i$ $B_i$ $w_i$ /$n$. 
In these expressions $N_i$ and $B_i$  are the number of events 
of the i$^{th}$ bin of the ``event map'' and ``background map'',
respectively,  
$w_i$ is a weight proportional to the value of the PSF 
at the angular distance of the i$^{th}$ bin, and
$n$ is the number of swaps.
The sum is over all the bins inside a radius R, chosen 
to contain the PSF.
Since the number of events per bin is large, 
the fluctuations follow the Gaussian
statistics, hence the errors on $N_{on}$ and $N_{off}$ are:
    $\delta N_{on}$ =  $\sqrt{\Sigma_i N_i w_i^2}$
and $\delta N_{off}$ = $\sqrt{\Sigma_i B_i w_i^2/n^2}$.

In order to study the signal in different energy regions,
the maps are built for 8 different N$_{pad}$ intervals, namely
20-39, 40-59, 60-99, 100-199, 200-299, 300-499, 500-999, and $>$1000.
These maps are then combined to have ``integral maps'' for
different  N$_{pad}$ thresholds.

Analysing the data recorded in 4 years, 
the sky maps of the  MGRO J1908+06 region
show a significant excess at the source position 
for different N$_{pad}$ thresholds.
The larger significance is given by events with  N$_{pad}\ge$ 20, 
with 7.3 standard
deviations. Increasing N$_{pad}$, the significance decreases. 
For N$_{pad}>$ 1000 no signal is present.

The distributions of the significances outside the source region follow
a standard normal distribution, showing the correctness of the 
background evaluation procedure.

As will be discussed in  Section 4.3, the signal
with N$_{pad}$=20-59 is largely affected by the 
Galactic diffuse gamma ray flux, and only events with
N$_{pad} \ge$ 60 will be used in the study of the source morphology and flux.
Fig.1 shows the significance map for events 
with N$_{pad} \ge$ 60, where the source signal
reaches 6.2 standard deviations.

Studying the source on the time scale of one year, the annual excess rate 
results to be consistent with the total average rate, indicating that
the gamma ray flux from MGRO J1908+06 is likely due
to a steady emission.

\subsection{Source position and extension}

To evaluate the position and extension of the source, the events
with N$_{pad} \ge$ 60 are used.
We assume a source shape 
described by a symmetrical two-dimensional 
Gaussian function with r.m.s. $\sigma_{ext}$.
Fitting the non-smoothed excess map 
to a function given by the convolution of the above Gaussian and the
detector PSF, we found the best-fit position at 
R.A. = 19$^h$08$^m$1$^s$ and 
decl. = 6$^{\circ}$24', with a statistical error of 12' and 
a systematic error of 6' per axis.
The position found is consistent with the the Milagro measurement
and with the centroid of HESS J1908+063
(R.A. = 19$^h$07$^m$54$^s$ and 
decl. = 6$^{\circ}$16'7'', with a statistical error of 2.4' and 
a systematic error of 20'' per axis).

The value of $\sigma_{ext}$ that best fits the data
is 0.49$^{\circ}\pm$0.22$^{\circ}$,
consistent with the H.E.S.S. estimation of 0.34$^{\circ}$.

Fig.2 shows the distribution of the 
angular distance from the best-fit centroid position,
compared with the simulated distributions corresponding to the extensions
 $\sigma_{ext}$=0.49$^{\circ}$ 
and $\sigma_{ext}$=0$^{\circ}$.
The two curves are normalized to the same number of excess events.

\begin{figure}
    \includegraphics[height=0.33\textheight]{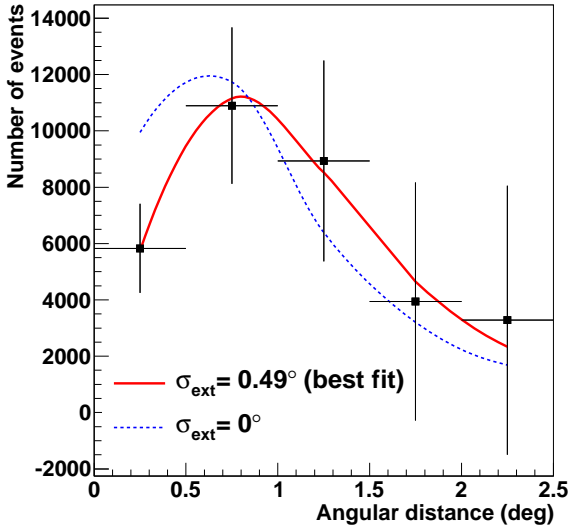}
    \caption{Number of excess events with N$_{pad} \ge$60 
as a function of the angular distance
from the best-fit centroid position, compared with the
expected distributions for different source extensions.
}
\vspace{15pt}
\end{figure}

\subsection{Energy spectrum}

In order to study the energy dependence of the signal 
the events are divided in different 
subsets according to the number of hit pads, 
and a sky map is built for each subset.
For this analysis we define 4 intervals:
N$_{pad}$ = 20-59, 60-199, 200-499 and N$_{pad} \ge$500.
The intervals have been chosen in order to have a signal of comparable 
statistical significance.

For the spectrum evaluation we assume a power law dependence: dN/dE = K E$^{-\gamma}$.
The values of K and $\gamma$ are derived 
by comparing the number of the excess events
detected in each of the previously defined N$_{pad}$ intervals
with the corresponding ones given by simulations assuming a set of test spectra.
The reliability of this procedure has been tested studying the 
the Crab Nebula signal \citep{Ver10}.

For each N$_{pad}$ interval, the number of excess events 
is obtained integrating the sky map around the source position
up to a distance $\psi_{max}$, where $\psi_{max}$ is the radius
of the opening angle which maximizes the signal to background ratio.
The value of $\psi_{max}$ depends on the source 
extension $\sigma_{ext}$ 
and on the detector PSF, and is provided by simulations.
For the extension we use the value $\sigma_{ext}$ = 0.49$^{\circ}$, 
according to our measurement. On the other hand, the PSF for a given N$_{pad}$ interval is not precisely determined, 
since it depends both on the detector characteristics and on
the spectrum index $\gamma$, which is unknown.
To solve this ``circular'' problem an iterative procedure has been applied. 

First, an initial index $\gamma$ = 2.5 is assumed, and the corresponding values of
$\psi_{max}$ for every N$_{pad}$ interval are determined via simulations.
The number of events observed in $\psi_{max}$ are then used to evaluate a new spectral slope
$\gamma$ which is returned to the first step, to calculate a new set of 
$\psi_{max}$, and so on.
Given the relatively weak dependence of the PSF 
on $\gamma$, a small number of iterations is sufficient to terminate 
the process successfully and provide the parameters of the best fit spectrum.

\subsection{Contribution from diffuse flux}

Since the source is located on the Galactic plane, the observed flux 
could be affected by the diffuse gamma ray emission
produced by cosmic rays interacting with the matter and 
the radiation fields of the Galaxy.
Given the relatively large opening angles used in the
measurement, the photons from the diffuse radiation
falling in the observational window of MGRO J1908+06 could artificially
increase the flux detected from the source direction.
The amount of this contribution can be evaluated by analysing the data 
collected from the Galactic plane region close to the source.

\begin{figure}
    \includegraphics[height=0.33\textheight]{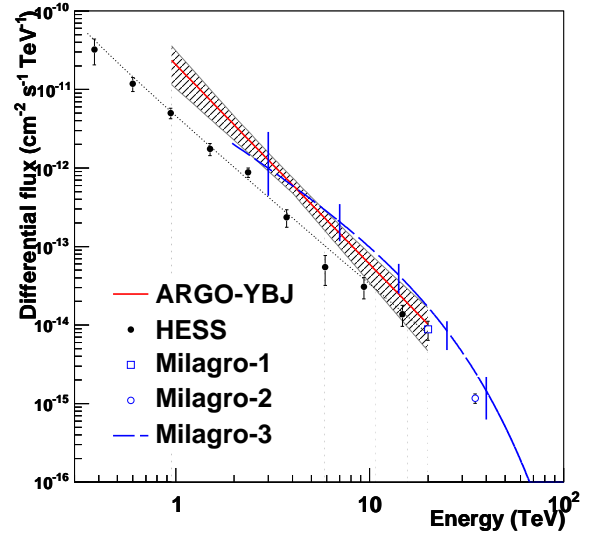}
    \caption{Gamma ray flux from MGRO J1908+06 measured by different detectors. 
ARGO-YBJ: the continuous line (colored red in the online version) 
is the best fit to data. The dashed area
represents the one standard deviation error.
H.E.S.S.: the dotted line is the best fit to the points \citep{hess09}.
Milagro 1: flux value assuming a spectrum $\propto$E$^{-2.3}$ \citep{mila07}.
Milagro 2: flux value assuming a spectrum $\propto$E$^{-2.6}$ \citep{mila09}.
Milagro 3: the dashed line (colored blue in the online version) 
is the spectrum fit according to Smith (2009) and the 
vertical lines are the errors (at one standard deviation) for some 
values of the energy. 
The plotted errors are purely statistical for all the detectors.}
\vspace{15pt}
\end{figure}

The flux of very high energy Galactic gamma rays
in the region of MGRO J1908+06 
(Galactic coordinates $l$ = 40.39$^{\circ}$ and $b$ = -0.79$^{\circ}$)
is poorly known.

The first evidence of a diffuse Galactic emission 
at TeV energies has been reported by the Milagro detector
\citep{mila05}.
A significant dependence of the flux 
on the Galactic latitude and
longitude has been found in a later analysis 
with events of median energy 15 TeV \citep{mila08}. 
The same paper reports the expected energy spectrum 
for two different sectors
of the Galactic plane, for energies from 10 keV to 100 TeV, 
according to the GALPROP model \citep{Stro00,Port08}, 
``optimized'' to fit the measurements by EGRET in the
40 MeV - 10 GeV energy range and by Milagro at 15 TeV.
Concerning the MGRO J1908+06 region, the expected average flux at 1 TeV 
in the area of Galactic coordinates
$l\in$ [30$^{\circ}$,65$^{\circ}$] and $b\in$ [-2$^{\circ}$,2$^{\circ}$] is 
$\sim$2$\times$10$^{-9}$ photons TeV$^{-1}$  cm$^{-2}$ s$^{-1}$ sr$^{-1}$.

A preliminary flux measurement at energies E $>$300 GeV, obtained with the
ARGO-YBJ data, is reported by Ma (2011), who derives the average 
gamma ray spectrum for $l\in$ [25$^{\circ}$,65$^{\circ}$] and
$b\in$ [-2$^{\circ}$,2$^{\circ}$]. 
This estimate results lower but still 
consistent with the expectation of the above model.

For our purposes, given the variation of the  
emission along the Galactic plane and its strong dependence on the latitude,
it is preferable to evaluate the diffuse flux
in a restricted region, adjacent to the source position.

We consider two sky regions, L$_1$ and L$_2$, of size 
$\Delta l$ = 5$^{\circ}$ and $\Delta b$ = 2$\times \psi_{max}$ 
(where  $\psi_{max}$ depends on the  N$_{pad}$ interval)
whose centers have the same latitude of MGRO J1908+06 and are located 
at both sides of the source, at a longitudinal distance 
of 5.5$^{\circ}$.

Analysing the ARGO-YBJ data, a global excess of statistical significance 3.0 
standard deviations 
over the cosmic ray background is observed 
in L$_1$ + L$_2$ for events with N$_{pad}>$20. 
This excess is interpreted as due to the diffuse Galactic emission 
plus the contribution of 5 gamma ray sources discovered
by H.E.S.S., namely HESS J1912+101 \citep{hess08b},
W49B \citep{hess11}, HESS J1857+026 and HESS J1858+020 \citep{hess08a}, 
and HESS J1849-000 \citep{hess08c}.

The individual fluxes of these objects 
are below the ARGO-YBJ sensitivity, while
the total flux is (29$\pm$3)$\%$ that of the Crab Nebula at 1 TeV \citep{ver11}.
In particular, HESS J1912+101 and HESS J1857+026
have a flux $\sim$10$\%$ and $\sim$17$\%$ that of the Crab Nebula,
respectively.
The number of events from these sources expected to fall 
in L$_1$ and L$_2$ is evaluated via simulations, using the fluxes
measured by H.E.S.S., and gives a global contribution of
(40$\pm$14)$\%$ to the observed diffuse excess.
After the subtraction of this contribution,
taking into account the different exposures of L$_1$ and L$_2$,
we evaluate the number of events due to the diffuse emission
expected to fall into the observational window of MGRO J1908+06.

We found that the ratio between the number of events  
expected from the diffuse emission 
and those observed from the source direction is
$R_d$ = 0.33$\pm$0.18 for showers with N$_{pad}$=20-59, and 
$R_d$ <0.15 (at one sigma level) for showers with N$_{pad}\ge$60. 

As a comparison, the values derived
by using the ``optimized'' GALPROP diffuse emission model 
given by Abdo et al.(2008) are
$R_d$ = 0.57 for N$_{pad}$=20-59, and 
$R_d$ = 0.23 for N$_{pad}\ge$60. 
These values are larger than 
those obtained with the ARGO-YBJ data.
It should be noted, however, that the above model is based on a measurement
by Milagro which does not take into account all the gamma ray sources 
located in the studied region, 
and could overestimate the diffuse flux.

The larger contribution of the diffuse emission
for N$_{pad}$=20-59
is due to the wider opening angle used in this interval 
($\psi_{max}$=2.0$^{\circ}$).
Because of these estimates, 
to avoid a possible large systematic effect in the flux evaluation, 
we restrict our spectral analysis to the events with N$_{pad}\ge$60.

Performing the procedure described in Section 4.2, we fit the data of the 
3 intervals N$_{pad}$=60-199, 200-499 and N$_{pad} \ge$500.
The best fit spectrum obtained is:
dN/dE = 6.1$\pm$1.4 $\times$ 10$^{-13}$ (E/4 TeV)$^{-2.54\pm0.36}$ 
photons cm$^{-2}$ s$^{-1}$ TeV$^{-1}$,
valid in the energy region 1-20 TeV.
The median energies corresponding to the 3 N$_{pad}$ 
intervals are 2.4, 5.1 and 12.8 TeV, respectively.

As a comparison, if we do not exclude the data with 
N$_{pad}$=20-59, the best-fit spectrum is:
dN/dE = 1.36$\pm$0.29 $\times$ 10$^{-12}$ (E/3 TeV)$^{-2.65\pm0.25}$ 
photons cm$^{-2}$ s$^{-1}$ TeV$^{-1}$,
which gives a flux 21$\%$ higher at E = 1 TeV.

Beside the statistical errors and the systematics due to the
diffuse contribution discussed before, 
our measurement could be affected by
an additional systematic error mainly due to the background evaluation,
to the absolute energy scale determination, 
to the pointing accuracy, to environmental effects and to the Monte Carlo
simulations,
for a global effect that we estimate to be $<$ 30$\%$ (Aielli et al. 2010).

In the case of an extended source, a possible further 
cause of systematics could be the uncertainty in the extension,
and the consequent use of an incorrect opening angle
to extract the signal.
Therefore we have 
also evaluated the spectrum assuming $\sigma_{ext}$ = 0.34$^{\circ}$, as 
measured by H.E.S.S. The resulting flux differs from the
previous one by less than 5$\%$ in the whole energy range 
considered in the analysis.

The obtained spectrum is shown in Fig.3, 
together with those reported by H.E.S.S. and Milagro.
The flux is significantly higher than that given by H.E.S.S. in the 1-10
TeV energy range, but is consistent with the Milagro spectrum
\citep{mila09b}.
The hard spectrum with exponential cutoff obtained by Milagro
produces a worse, but still acceptable, fit to our data.
Given the reduced significance 
of the excess at high energies, we are not able to constrain the shape of the
spectrum above 10 TeV and to definitively rule out a high energy cutoff.

\section{Discussion and conclusions}

The gamma ray source MGRO J1908+06 has been studied by ARGO-YBJ
analyzing $\sim$4 years of data. 
An excess with significance 6.2 standard deviations
is observed in a position consistent
with previous measurements by Milagro and H.E.S.S..
The peak of the signal
occurs at R.A. = 19$^h$08$^m$1$^s$ and 
decl. = 6$^{\circ}$24'
(with statistical and systematic errors of $\sim$0.2$^{\circ}$ and  
0.1$^{\circ}$ per axis, respectively)
and lies at a distance of 22' from PSR J1907+0602, 
consistent with the pulsar location within the measurement error.

The signal is due to emission from an extended region. After taking into
account the detector PSF, the extension of the source is found to be
$\sigma_{ext}$= 0.49$^{\circ}$$\pm$0.22$^{\circ}$.

The photon spectrum in the range 1-20 TeV follows a simple power law
with a spectral index 2.54$\pm$0.36, though a harder spectrum with a
high energy cutoff cannot be ruled out.

The spectrum is found to be consistent with the Milagro result \citep{mila09b}
but not with the H.E.S.S. best fit in the 1-10 energy range,
the flux measured by ARGO-YBJ at 4 TeV being a factor 2.6 larger.
At $\sim$ 20 TeV the ARGO-YBJ, H.E.S.S. and Milagro fluxes are consistent
within the errors,
and are also in agreement with the first Milagro measurement \citep{mila07}.

Since a contribution to this measurement is expected from the Galactic
diffuse emission, data from two sky regions located at both sides of the
source and centered at the same latitude have been used to determine this
contamination.
According to this estimate, the diffuse Galactic gamma ray emission
is expected to contribute to the signal above 1 TeV for less than
$\sim$ 15$\%$, and cannot account for the observed disagreement.

Being the difference with H.E.S.S. at the level of 2.5 standard deviations, 
the discrepancy
could be simply due to statistical fluctuations, or to the combination
of statistical and systematic uncertainties. 
However these latter have been accurately studied, giving 
a global error on the flux less than 30$\%$.
Indeed, the spectrum of the Crab Nebula obtained by ARGO-YBJ results
to be in good agreement with the Cherenkov detectors measurements \citep{Ver10,ver11}.
The extension of the source should not give 
such an additional systematic error to explain the observed difference.

On the other hand, a similar discrepancy is found in the observation of the 
extended source MGRO J2031+41, located in the Cygnus region, 
for which ARGO-YBJ \citep{Bar12} and Milagro \citep{mila12} report
a flux significantly larger than that measured by the Cherenkov Telescopes
MAGIC and HEGRA.

In principle, one cannot exclude  
the possibility of a flux variation as the origin 
of the observed disagreement among the detectors. 
Milagro, H.E.S.S. and ARGO-YBJ data 
have been recorded in different periods. 
Milagro integrates over seven years (July 2000 - November 2007) 
while the total H.E.S.S. data set only amounts to 27 hours of
sparse observations during 2005-2007, before the ARGO-YBJ measurement.
However, a possible flux variation seems unlikely, since
the average fluxes measured by Milagro and ARGO-YBJ in two contiguous 
periods covering a total time of 11 years, are consistent.

Moreover, it should be noted that if MGRO J1908+06 
is the pulsar wind nebula associated to PSR J1907+0602,
the gamma ray emission
originates from a region whose size has been estimated 
to be $\ge$40 pc \citep{fermi10}, implying that 
the variation time scale cannot be less than $\sim$130 years,
unless relativistic beaming effects are present.

In conclusion, MGRO J1908+06 is observed by ARGO-YBJ as a stable extended
source, likely the TeV nebula of PSR J1907+0602, with a flux at 1 TeV
$\sim$67$\%$ that of the Crab Nebula. Assuming a distance of 3.2 kpc,
the integrated luminosity above 1 TeV is $\sim$1.8 times 
that of the Crab Nebula,
making  MGRO J1908+06 one of the most luminous Galactic gamma ray sources
at TeV energies.

\acknowledgments

This work is supported in China by NSFC (No. 10120130794), the Chinese Ministry
of Science and Technology, the Chinese Academy of Science, the Key Laboratory
of Particle Astrophysics, CAS, and in Italy by the Istituto Nazionale di
Fisica Nucleare (INFN).

We also acknowledge the essential support of W.Y. Chen, G. Yang, X.F. Yuan, 
C.Y. Zhao, R. Assiro, B. Biondo, S. Bricola, F. Budano, A. Corvaglia,
B. D'Aquino, R. Esposito, A. Innocente, A. Mangano, E. Pastori, C. Pinto,
E. Reali, F. Taurino, and A. Zerbini, in the installation, debugging, and
maintenance of the detector.




\begin{thebibliography}{99}


\bibitem[Abdo et al.(2007)]{mila07} Abdo, A.A., Allen, B., Berley, D., et al. 2007, ApJ, 664, L91
\bibitem[Abdo et al., 2008]{mila08} Abdo, A.A., Allen, B., Aune, T., et al. 2008, ApJL 688, 1078
\bibitem[Abdo et al., 2009a]{ferm09} Abdo, A.A., Ackermann, M., Ajello, M., et al. 2009a, ApJS 183, 46
\bibitem[Abdo et al., 2009b]{mila09} Abdo, A.A., Allen, B.T., Aune, T., et al. 2009b, ApJ, 700, L27
\bibitem[Abdo et al., 2010]{fermi10} Abdo, A.A., Ackermann, M., Ajello, M., et al. 2010, ApJ 711, 64
\bibitem[Abdo et al., 2012]{mila12} Abdo, A.A., Abeysekara, U.,  Allen, B.T., et al. 2012, ApJ, 753, 159
\bibitem[Aharonian et al., 2008a]{hess08a} Aharonian, F., Akhperjanian, A.G., Barres de Almeida, U., et al. 2008a, A\&A, 477, 353
\bibitem[Aharonian et al., 2008b]{hess08b} Aharonian, F., Akhperjanian, A.G., Barres de Almeida, U., et al. 2008b, A\&A, 484, 435
\bibitem[Aharonian et al., 2009]{hess09} Aharonian, F., Akhperjanian, A.G., Anton, G., et al. 2009, A\&A, 499, 723
\bibitem[Aielli et al., 2006]{Aie06} 
Aielli, G., Assiro, R., Bacci, C., et al. 2006, Nucl. Instrum. Methods Phys. Res. A, 562, 92
\bibitem[Aielli et al., 2008]{Aie08} 
Aielli, G., Bacci, C., Barone, F., et al. 2008, Astrop. Phys., 30, 85
\bibitem[Aielli et al., 2009]{Aie09} Aielli, G., Bacci, C., Bartoli, B., et al. 2009, Astrop. Phys., 30, 287
\bibitem[Aielli et al., 2010]{Ver10} Aielli, G., Bacci, C., Bartoli, B., et al. 2010, ApJ, 714, L208
\bibitem[Alexandreas et al., 1993]{Ale92} Alexandreas, D.E., Berley, D., Biller, S., et al. 1993, Nucl. Instrum. Methods Phys. Res. A, 328, 570
\bibitem[Aloisio et al., 2004]{Alo04} Aloisio, A., Branchini, P., Catalanotti, S. et al. 2004, IEEE Transaction on Nuclear Science, 51, 1835
\bibitem[Amenomori et al., 2010]{Ane10} Amenomori, M., Bi, X.J., Chen, D., et al. 2010, ApJ, 709, L6
\bibitem[Atkins et al., 2005]{mila05} Atkins, R., Benbow, W., Berley, D., et al. 2005, Phys. Rev. Lett. 95, 251103
\bibitem[Bartoli et al., 2011]{DiS11} Bartoli, B., Bernardini, P., Bi, X.J., et al. 2011, Phys. Rev. D, 84, 022003
\bibitem[Bartoli et al., 2012]{Bar12} Bartoli, B., Bernardini, P., Bi, X.J., et al. 2012, ApJ, 745, L22
\bibitem[Brun et al., 2011]{hess11}
Brun, F., De Naurois, M., Hofmann, W., et al. 2011, in Proc. 25th Texas Symposium on Relativistic Astrophysics (available at arXiv:1104.5003v1)
\bibitem[Fleysher et al., 2004]{Fle04} Fleysher, R., Fleysher, L., Nemethy, P., \& Mincer, A.I. 2004, ApJ, 603, 355
\bibitem[GEANT, 1993]{Gea93} 
GEANT - Detector Description and Simulation Tool  1993, CERN Program 
Library, W5013, http://wwwasd.web.cern.ch/wwwasd/geant/
\bibitem[Heck et al., 1998]{Hec98} 
Heck, D., Knapp, J., Capdevielle, J.N., Shatz, G., \& Thouw, T. 1998, Forschungszentrum Karlsruhe Report No. FZKA 6019
\bibitem[He et al., 2007]{He07}
He, H.H., Bernardini, P., Calabrese Melcarne, A.K., Chen, S.Z. 2007, Astropart. Physics, 27, 528
\bibitem[Iacovacci et al., 2009]{iaco09} Iacovacci, M., Corvaglia, A., Creti, P., et al. 2009, in Proc. 31st Int. Cosmic Ray Conf., Lodz, Poland (available at http://icrc2009.uni.lodz.pl/proc/html/)
\bibitem[Ma, 2011]{Ma11} Ma, L.L. 2011, in Proc. 32nd Int. Cosmic Ray Conf., Beijing, China (available at http://www.ihep.ac.cn/english/conference/icrc2011/paper/)
\bibitem[Nolan et al., 2011]{fermicat} Nolan, P.L., Abdo, A.A., Ackermann, M., et al. 2011, ApJS, in press (arXiv: 1108.1435)
\bibitem[Porter et al., 2008]{Port08} Porter, T.A., Moskalenko, I.V., Strong, A.W., Orlando, E., \& Bouchet, L. 2008, ApJ 682, 400 
\bibitem[Smith 2009]{mila09b} Smith, A.J. 2009, in Proc. Fermi Symposium, Conf Proceedings C091122, http://www.slac.stanford.edu/econf/C0911022/
\bibitem[Strong et al., 2000]{Stro00} Strong, A.W., Moskalenko, I.V., Reimer, O. 2000, ApJ, 537, 763
\bibitem[Terrier et al., 2008]{hess08c}
Terrier, R., Mattana, F., Djannati-Atai, A., et al 2008, in Proc. 4th Int. Meeting on High Energy Gamma-Ray Astronomy, AIP Conference Proceedings, 1085, 312
\bibitem[Vernetto, 2011]{ver11}
Vernetto, S. 2011, in Proc. 32nd Int. Cosmic Ray Conf., Beijing, China (available at //www.ihep.ac.cn/english/conference/icrc2011/paper/)
\bibitem[Ward, 2008]{ward08} Ward, J.E. 2008, in AIP Conf. Proc. 1085,
High energy Gamma-Ray Astronomy, ed. F.A.Aharonian, W.Hofmann, \& F.Rieger (Melville, NY:AIP), 301
\bibitem[Zhang, 2003]{Zha03} 
Zhang, J.L. 2003, in Proc. 28th Int. Cosmic Ray Conf., Vol.4, ed. T.Kajita et al. (Tokyo: Universal Academy Press, Inc.), 2405

\end{thebibliography}
\end{document}